# Investigating Taxi and Uber Competition In New York City: Multi-Agent Simulation by Reinforcement-Learning

Saeed Vasebi[a], Yeganeh M. Hayeri[ab]

[a] *Stevens Institute of Technology, 1 Castle Point, Hoboken, NJ 07030, USA*
[b] *Corresponding Author*

**Abstract**

The taxi business has been overly regulated for many decades. Regulations are supposed to ensure safety and fairness within a controlled competitive environment. By influencing both drivers' and riders' choices and behaviors, emerging e-hailing services (e.g., Uber and Lyft) have been reshaping the existing competition in the last few years. This study investigates the existing competition between the mainstream hailing services (i.e., Yellow and Green Cabs) and e-hailing services (i.e., Uber) in New York City. Their competition is investigated in terms of market segmentation, emerging demands, and regulations. Data visualization techniques are employed to find existing and new patterns in travel behavior. For this study, we developed a multi-agent model and applied reinforcement learning techniques to imitate drivers' behaviors. The model is verified by the patterns recognized in our data visualization results. The model is then used to evaluate multiple new regulations and competition scenarios. Results of our study illustrate that e-hailers dominate low-travel-density areas (e.g., residential areas), and that e-hailers quickly identify and respond to change in travel demand. This leads to diminishing market size for hailers. Furthermore, our results confirm the indirect impact of Green Cabs' regulations on the existing competition. This investigation, along with our proposed scenarios, can aid policymakers and authorities in designing policies that could effectively address demand while assuring a healthy competition for the hailing and e-haling sectors.

## 1. Introduction

Yellow Cab is an iconic feature of New York City (NYC), making over 146 million trips per year[1]. Yellow Cab was the only main hailing service in the NYC area for decades; however, the situation has significantly changed over the last few years as Green Cab and Uber have been introduced. In 2010, Uber started its e-hailing services in NYC, and now it makes about 100 million trips annually[2]. The local government known as NYC Taxi & Limousine Commission (TLC) introduced Green Cab in 2013 to enhance hailing services in NYC and especially in Outer Boroughs (i.e., Brooklyn, Queens, and Bronx) and to address unmet demand. Green Cab provides hailing services with about 19 million annual trips[1].

Today, Yellow Cab, Green Cab, and Uber are major players of the hailing and e-hailing services in NYC, addressing most of the taxi demand. Competition between these players influences their behaviors resulting in market segmentations[3]. Technological, regulatory, and policy advantages and disadvantages play significant roles in this competition and its consequential market segmentation. For instance, Uber's technological advancements, utilizing smartphones and communication technologies, impacts passengers' and drivers' lifestyles resulting in efficient access to e-hailing services. On the other hand, Green Cab's competition with Yellow Cab is limited by geographical regulations. Table 1 illustrates the differences in geographical regulations imposed on each group. Yellow Cab can pick up passengers at any location while Green Cab has a limited operation area. A TLC study

Corresponding author's contact infromation: yhayeri@stevens.edu, 525 River Street, Hoboken, NJ 07030

shows that 95% of Yellow Cab pickups are in Manhattan below 96th Street as well as JFK and LaGuardia airport[4]. This led to a regulatory decision made by TLC to limit Green Cab's pickup area to outside of core Manhattan, JFK, and LaGuardia airports. Uber has no geographical limitation except for not being able to pick-up passengers at the terminals, rather in dedicated airport lots.

Table 1- Taxis' geographical regulations and hailing technologies

| Taxi Provider | Introduction Date | Hailing /E-Hailing | Passenger Pick-Up Area |
|---|---|---|---|
| Yellow Cab | 1907 | Haling | Any location in NYC |
| Green Cab | August 2013 | Haling | Any location in NYC except below West 110th and East 96th streets in Manhattan, JFK, and LaGuardia Airports |
| Uber | May 2011 | E-hailing | Any location in NYC |

In this study, we provide insights into the existing competition between hailing and e-haling services with respect to market segmentation, geographical regulation impacts, and respond to new demands. We attempt to answer the following questions:
- How are hailing and e-hailing services' market segments different?
- How do hailing and e-hailing services respond to new demands or market expansion?
- How do Green Cab's geographical regulations influence Uber and Yellow Cab's operation?

In the following section, existing research in hailing and e-hailing services is reviewed. Then, we present the method and data used for this study. The result section provides insights into the taxis' behaviors with a multi-agent model and simulation. Finally, we present the outcome of the study and recommendations for future works within this realm. The contribution of this paper is to provide a transportation agent-based mobility model with learning capability to enhance the operation of taxis. The model has a real-world application that has been evaluated for NYC taxis and their imposed regulations. The reinforcement learning capability coded into the model is a valuable method to address dynamic changes in drivers' behaviors.

## 2. Literature review

The behavior of Taxi drivers has been widely studied over the last few years [2,5–10]. Liu et al. investigate taxis behavior in Shanghai, China, to find a relationship between taxis' drop-off and pick-up locations and land use[5]. They clustered the city based on different land uses and taxi pick-up and drop off locations. They found a relationship between the number of pick-up/drop-off and land use types (e.g., residential, commercial, and industrial) as well as land-use intensity. Austin et al. used Boston's taxicabs' dataset to study taxi distribution in relationship with population, land use, and transit accessibility[6]. Results showed that taxi distribution in Boston does not correlate with users' demand, and there is a significant gap between passenger and taxi distribution. Haggag et al. studied Cabs' behavior in NYC to find the best patterns for the drivers to find more passengers and earn more money[7]. Dong et al. analyzed the behavior of top-earning taxi drivers in NYC and they visualized movements of this group[8]. Their study revealed that top-earning drivers choose specific short trips and have better routing information.

E-hailing services have significantly altered hailing services as well as taxi drivers' behaviors [2,9–12]. Geradin investigated the competition of Uber and conventional taxis in Europe[10]. He concluded that e-hailing technology is an unavoidable revolution, and conventional hailing companies should accept this revolution and attempt to improve their regulations and technologies accordingly. Sun and Posen have also reported Uber's influence, and success in e-hailing services is a result of conventional taxis' outdated regulations, which stopped providing flexible, competitive, and convenient services for decades[11,12]. A recent investigation shows that competition between Uber and cab drivers in NYC and Chicago has reduced passengers' official complains about cab drivers[13]. Moreover, Cramer et al. compared the capacity utilization rate of UberX and conventional taxis in five cities of Boston, Los Angeles, NYC, San Francisco, and Seattle[9]. They found that UberX has higher capacity utilization than taxis. A comprehensive data visualization of Yellow Cab, Green Cab, and Uber's trips in NYC has been performed by Toddwschneider website[2]. It has visualized increasing trends of Uber and Green Cab trips from 2014 and decreasing trips for Yellow Cab. To our knowledge to date, detailed influence of e-hailing and hailing services on the market, regulations' impact, and market expansion has not been investigated. In the next sections, we utilize qualitative and quantitative methods to provide insights on these topics.

## 3. Method and data

In this study, we apply data visualizations and multi-agent modeling with learning capability to explore behaviors of haling and e-hailing drivers. Our approach is shown in Figure 1. First, we visualize the geographic data of Yellow Cab, Green Cab, and Uber. Visualizations are concentrated on e-hailing and hailing differences and regulations' impacts. The visualizations are used to find patterns in behaviors of the taxis. Based on the identified patterns, a set of hypotheses are proposed. A multi-agent model has been developed to investigate these hypotheses and validate the model. We then use the model to investigate a set of new scenarios for the future of hailing and e-hailing services.

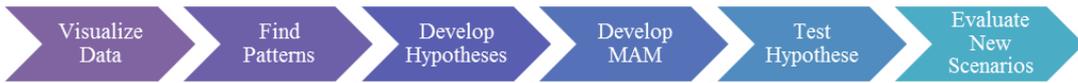

Figure 1 - Study's Approach

In order to simulate the realistic behavior of the agents, Yellow Cab, Green Cab, and Uber, a multi-agent model with learning capability has been coded. This model utilizes Q-learning method to imitate drivers' behavior when they are searching for passengers. Q-learning is a Reinforcement Learning technique introduced by Watkins[14]. In this technique, each agent has a set of states (*S*), in this case, the location of the agents. In each state; the agent could perform a set of actions (*A*) (i.e., the potential directions for the move). The agent moves from one state to another state by performing an action. The agent receives a reward (*r*) in each state. Picking passengers are considered as rewards. The goal of the model is to maximize the agent's total reward by performing optimum actions. Optimum action for each state is estimated by the total expected value of rewards of that action. Quality of state-action is started from zero and updated each episode as:

$$Q(s_t, a_t) \leftarrow (1-\mu).Q(s_t, a_t) + \mu.\left(r_t + \gamma.\max_a Q(s_{t+1}, a)\right)$$

Where $Q(s_t, a_t)$ is the quality of action $a$ at state $s$ in time $t$, $\mu$ is learning rate of the model. Zero means no learning in this step and 1 means only learn from this step and forget previous learning. $r_t$ is a reward in this state, $\gamma$ is discount value, showing how current reward is important in comparison with future potential rewards, and $\max_a Q(s_{t+1}, a)$ is quality of optimum action in the next step. For detailed information on Q-learning you may refer to Cybenko [15].

We coded the multi-agent model in NetLogo software. The model considers Cabs and Uber as agents looking for passengers. The key difference between Uber and Cabs is that Cabs should be in the same block as passengers to pick them up, but Uber is able to pick them up from one block away too. This assumption has obtained from technological capability of Uber, which utilizes an app to remotely find passengers, versus a cab driver who should be close enough to passengers to see them. For more detailed information, you may refer to Appendix A.

Dataset for this study is captured from TLC's historical database, which includes more than one billion taxi trips since January 2009[16]. Additionally, we used over 19 million Uber trips data from April to September 2014[17].

## 4. Results

### 4.1. Data visualization

We used data visualizations to find drivers' behavioral patterns in each group of taxis. Trips are visualized based on longitude and latitude of pick-up locations from April to September 2014. Drop-off locations are not considered in this study since drivers do not have control of drop-off locations.

A broad view of the taxis' pick-up locations indicates each group of taxis has different operation areas (see Figure 4 in Appendix B). Yellow Cab mainly concentrates in Manhattan, JFK, and LaGuardia airports, where there is a high-density of potential taxi passengers. A few Yellow Cab's pickups occur in Outer Boroughs, including Brooklyn, Queens, and the Bronx. In contrast, Green Cab mostly picks up passengers in Outer Boroughs and in its operation area in Upper Manhattan. Green Cab focuses on low passenger density areas such as Brooklyn and Queens since the regulations do not allow Green Cabs to pick up passengers in high-density areas such as the airports and Lower Manhattan. Uber is the only taxi service that is active in all NYC Boroughs. While Manhattan is the main market for Uber, it picks up many passengers in Queens and Brooklyn, which have a lower density of potential passengers compared with other areas.

Detailed analyses of passenger pickups in Queens and Brooklyn illustrates different behavior of hailing and e-hailing services (see Figure 5 in Appendix B). Yellow Cab has a few pickups in Outer Borough, and these pick-ups happen around transportation hubs (e.g., bus terminals and subway stations). Green Cab also picks up passengers around transportation hubs (not the terminals themselves) and main streets where it has a higher chance to find passengers. However, Uber finds its passengers in both transportation hubs and residential areas (e.g., in front of passengers' homes).

Data visualization of taxis from April to September 2014 shows the market expansion of Uber is different from Cabs in this period (see Figure 8 and detailed descriptions in Appendix B). While Cabs are mainly concentrated on traditional market for passengers (e.g., subway stations, bus terminal, and stadiums), Uber has quickly expanded pick-ups in new markets (e.g., museum, theatre, and parks), where there has been a nonexistent market for hailing services before.

Another pattern is identified around West $110^{th}$ and East $96^{th}$ streets in Manhattan, the border of Green Cab operation area (see Figure 6 in Appendix B). Due to the regulations, Green Cab is not allowed to pick-up any passenger below these streets. Our visualizations illustrate a drastic reduction in Yellow Cab and Uber pick-ups above these streets. Yellow Cabs and Ubers are allowed to pick-up passengers below and above these streets, but their pick-ups significantly decrease above the streets where Green Cab's operation area starts.

Our visualizations indicate Green Cab has a considerable number of pickups around LaGuardia airport (e.g., Jackson Heights neighborhood) in comparison with Yellow Cab and Uber (see Figure 7 in Appendix B). Yellow Cab and Uber pick up a large number of passengers at the airport terminals and dedicated lots. The regulation does not allow Green Cab to pick up passengers at the airport, and Green Cab could only drop off passengers. However, Green Cab's number of pickup around LaGuardia airport (i.e., except the airport itself) is much higher than Uber and Yellow Cab.

*4.2. Hypotheses*

Using the result of data visualizations and the patterns identified, the following hypotheses are proposed to set and validate our multi-agent model and simulation:

A. <u>Hailing and e-hailing services' market segmentations are different</u>: Cabs need face-to-face contact to find passengers; therefore, they only concentrate on high passenger density areas to have a higher chance for passenger pick-up. Uber is successful to pick-up passengers at both high and low-density passenger areas.

B. <u>Hailing and e-hailing services expand their markets and respond to new demands differently</u>: Uber is more successful in identifying and responding to new demands or changes in demand for travel. Due to availability of smart technologies, when a new market is emerging, or demand for taxi is changing, Uber can adapt quickly, while it would take some time for the Cab drivers to learn these changes on their own.

C. <u>Green Cab's regulations have indirect impacts on Uber and Yellow Cab at operation border area</u>: Green Cabs dropping off their passengers inside Manhattan should drive up to West $110^{th}$ and East $96^{th}$ streets for next pick-up; therefore, the number of Green Cabs above these streets has increased while Yellow Cab and Uber drivers prefer to not hail above these streets. As a result, Yellow Cab and Uber's number of pick-up drastically drops above these streets. Also, passengers, who could easily find Green Cab above the streets, prefer to hail a Cab, which could be faster and easier than e-hailing an Uber.

D. <u>Green Cab's regulations have indirect impacts on Uber and Yellow Cab at the LaGuardia airport</u>: Green Cabs dropping off passengers at the airport are not allowed to pick up passengers in the terminal. They drive to the neighboring areas to pick up passengers. Thus, Green Cab's number of pick-ups is relatively high in surrounding areas. Yellow Cabs and Uber, allowed to pick up passengers at the airport, prefer to not hail in neighboring areas. As a result, Green Cab's number of pick-ups goes even higher than Uber and Yellow Cab around the airport.

*4.3. Evaluation of the hypotheses with the multi-agent model*

Based on the proposed hypotheses, a multi-agent model has been configured. Details of the multi-agent configuration, the hypotheses modeling, and detailed results are provided in the appendix. Results are as follows:

*Hypothesis A* investigates how Uber is more successful than Cabs in low-density passenger areas. The results show Uber picks up 34% of total passengers in high-density areas and 42% of total passengers in low-density areas. Cabs picked up 14% and 10% of total passengers respectively in high and low-density areas. These results are aligned with *Hypothesis A* that Cabs have a higher pick-up performance in high passenger density areas than low-density areas, and Uber has a relatively better performance in the low-density areas.

The simulation results for *Hypothesis B* show Uber's market share from new demand is higher than its market share from fixed demand. Uber picks up 76% of fixed demand, and when new demand is generated in the simulation, Uber picks up 81% of the new demand. Conversely, Cabs pick up 24% of fixed demand, but it could only gain 19% of new demand in the simulation. Therefore, Cabs' market share from new demand is lower than their market share from fixed demand. This result is aligned with *Hypothesis B*.

Our simulation for *Hypothesis C* demonstrates when Green Cab is introduced to the model, Yellow Cab and Uber's numbers of pick-ups in the Green Cab operation area are decreasing by 30% and 16%, respectively. These results support *Hypothesis C*. Also, the results show that Yellow Cab is more affected by Green Cab's regulation than Uber.

*Hypothesis D* claims since Green Cab is prohibited to pick-up passengers at the LaGuardia airport, Green Cab's pick-ups around the airport are much higher than other taxis. The result shows that Yellow Cab's pick-up in the Green Cab operation area, surrounding neighborhoods, is 53% lower than the Green Cab prohibited area, and Uber has 9% fewer pick-ups. So, this result clarifies Yellow Cab and Uber's pick-ups are influenced by Green Cab's regulation.

*4.4. Development of scenarios*

Based on the identified patterns and evaluated hypotheses, we develop and evaluate two scenarios for the future of NYC taxis. The scenarios are:

A. Uber Ban: In this case, TLC might conclude that conventional Cabs could not compete with Uber and propose a regulation to limit the operation area of Uber. This scenario keeps Uber out of the Core Manhattan area. This scenario would not impose any regulation on Yellow Cab.

B. App for all taxis: Since app technology is Uber's main advantage to find passengers quickly and efficiently, TLC might apply the same technology for Green and Yellow Cabs. So, Cabs could pick up passengers by hailing and e-hailing. In this scenario, all hailing and e-hailing groups would have the same technological advantage.

These scenarios have been coded and run in our multi-agent model. The results illustrate that if the operation area of Uber is limited, Manhattan will be an exclusive market for Yellow Cab. So, Yellow Cab's pick-up drastically increases by 330%. Also, Uber will then gain a market share outside of Manhattan, and its pick-ups in Outer Borough will increase by 13%. This regulation significantly will influence Green Cab operation, which would lose 59% of its market.

In a scenario where all taxis are equipped with an e-hailing app, Uber and Yellow Cab, each will own 44.5% of total pick-ups (i.e., 91% increase in Yellow Cab percent of pick-up) and Green Cab will share 11% of the market (i.e., 41% increase). Therefore, hailing and e-hailing services will reach a healthier level of competition.

**5. Discussion and conclusion**

This study has investigated competition between hailing and e-haling services with regards to market segmentations and regulations' impacts in NYC. We visualized pick-up locations of taxis in NYC and identified differences in Cabs (i.e., Yellow and Green Cabs) and Uber's drivers' behaviors. A set of hypotheses and scenarios have been proposed and evaluated by the multi-agent simulation with learning capability. Our results illustrate that:

E-hailing taxis are more successful than hailing taxis to find passengers in low-density passenger areas. E-hailing taxis gain this advantage due to their app technology, which aid drivers to remotely find passengers; while hailing taxis should be close enough to the passengers to see them. So, hailers have a relatively low chance of passenger pick-up in low passenger density areas. Thus, they drive to high-density areas (e.g., transportation hubs, subway stations, and central business districts), where they could find a more efficient market.

E-hailers could rapidly identify the change in demand (i.e., emerging market, seasonal changes, or temporary demand) compared with hailers. E-hailing technology creates a network of passengers and drivers, which informs e-hailers of new/unmet demands in different ways. However, hailing drivers should search for passengers by themselves, and although they eventually identify changes in demand, the process is more gradual. Additionally, the e-hailing service database provides an opportunity for the company to analyze data in a systematic manner and to provide insights to drivers as needed to cope with changes in demand. These insights result from a collective learning method (i.e., learn in a network of drivers), hence they respond to market demand more efficiently. To take advantage of these opportunities, it is strongly recommended that hailing companies improve their data analytics capabilities to look into not only demand but factors such as seasonal changes, efficient pick-up areas, and emerging markets.

Indirect impacts of Green Cab's geographical regulations on Yellow Cab and Uber's operation are demonstrated in this study. Our results show when regulations prevent competition of the taxis in an area, the regulated service concentrates on the underserved area resulting in market share loss for other less regulated services. The behavior of taxi drivers is dynamic, and it finds a new way to react to new regulations.

We used a multi-agent simulation to evaluate a set of potential scenarios for the future of taxi services. These scenarios, and potentially more in the future, can help agencies to see the impact of their proposed regulations on other services and the system as a whole. For instance, imposing regulations on Uber can drastically and negatively impact the Green Cabs operation and market share. In another scenario, we examined the possibility of providing a technology app for all hailing services, which is what NYC TLC has been working on since 2015.[1] Such investment can increase the market share of both hailing services in the City.

This study has a few limitations and future work opportunities. First, the Uber dataset utilized in this study is from 2014. This is the only Uber's accurate dataset available. The analyses and their results could be revisited as an updated dataset becomes available. This model could be improved by considering passengers' preference between hailers and e-hailer (e.g., price, convenience, and wait time). Besides, our multi-agent model can be utilized to study taxis behaviors in other cities or to evaluate new scenarios as deemed appropriate. The goal of analyzing scenarios is to provide quantitative analysis, insights, and recommendations to the local government and policymakers before proposing regulations and to minimize the negative impact of regulations by providing a holistic view of the proposed and new laws and regulations.

## Appendix A. Multi-agent modeling

We used a multi-agent model which considers Cabs (i.e., Yellow and Green Cabs) and Ubers as agents. The agent moves blocks one-by-one, and they look for passengers (see Figure 2). The key difference between Uber and Cabs is that Cabs should be in the same block as passengers to pick them up, but Uber is able to pick them up from one block away too. This capability is shown as the blue rings around Ubers in Figure 2. This assumption has been obtained from the technological capability of Uber, which utilizes an app to find customers, versus Cabs, which should be close enough to passengers to see them.

Passengers in each block are generated by a random function. The areas with a high-density of passengers (e.g., Manhattan, and transportation hubs) are simulated by a passenger generation function with a higher probability of passenger generation than low passenger density areas (e.g., residential areas and Queens). In Figure 2, low and high passenger density areas are shown in yellow and red colors, respectively.

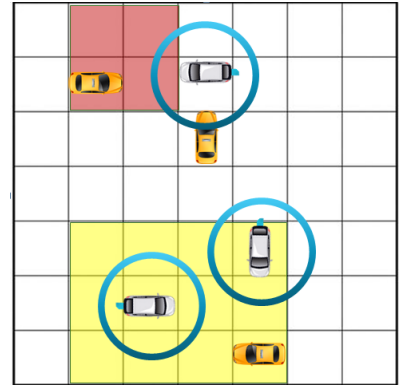

Figure 2- The multi agent model's conceptual view

To reflect the realistic behavior of Uber and Cabs, Q-Learning reinforcement method has been employed to model looking for passenger behavior. Based on Q-Learning method, drivers learn to choose the best direction to move, which brings them faster to a reward (i.e., passenger pickup). When a taxi finds a passenger, it drops off the passenger in a random location and starts looking for the next passenger. The algorithm of learning for this study is shown in Figure 3. For model initiation, the gridded world (i.e., simulation area) is generated, passenger generation functions are assigned, and taxis are randomly distributed in the gridded world. In each iteration, new passengers are generated. Then the taxis move one block. If they meet a passenger, they will pick it up and drop off in a random location. If taxis do not meet any passenger, they will wait for the next iteration. At the end of each iteration, the taxis' learning function is updated. The model has been coded, and its variables are calibrated to reach a satisfactory learning level.

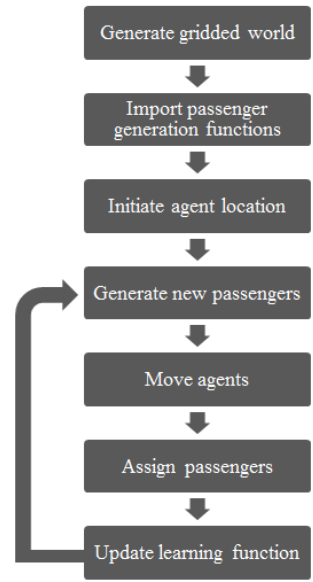

## Appendix B. Data visualizations, model configurations, and results

In the visualizations, since Yellow Cab's number of pick-ups is much larger than others (i.e., one order of magnitude), we used the same size samples to highlight differences in the distribution of taxis' pick-ups. A sample of 300,000 trips in a month has been selected for each group of taxis. Trips are visualized based on longitude and latitude of pick-up locations in each month. Figure 4 shows a broad view of Yellow Cab, Green Cab, and Uber's pick-ups during September 2014. The picture on the right shows Yellow Cab pick-ups, and the picture on the left shows Uber and Green Cab's pick-ups. Pick up the location of Yellow Cab, Green Cab, and Uber are shown by yellow, green, and red nodes, respectively.

Figure 3- Algorithm of the multi agent model

Figure 5 shows passengers' pick-ups in Park Slop neighborhood in Brooklyn, which represents drivers' behavior in most parts of Outer Boroughs. There are very few Yellow Cab pick-ups in these areas, and the pick-ups occur around transportation hubs and subway stations. Green Cab covers most of the pick-ups in main streets, transportation hubs, and subway stations. In contrast, Uber picks up passengers in small streets and residential areas.

Figure 6 shows around West 110[th] and East 96[th] streets in Manhattan, where is the border of Green Cab operation area. Green Cab is not allowed to pick-up any passengers below these lines. So, there is no green node below these streets. However, despite no regulations, Yellow Cab and Uber's pick-ups are drastically reduced above this border.

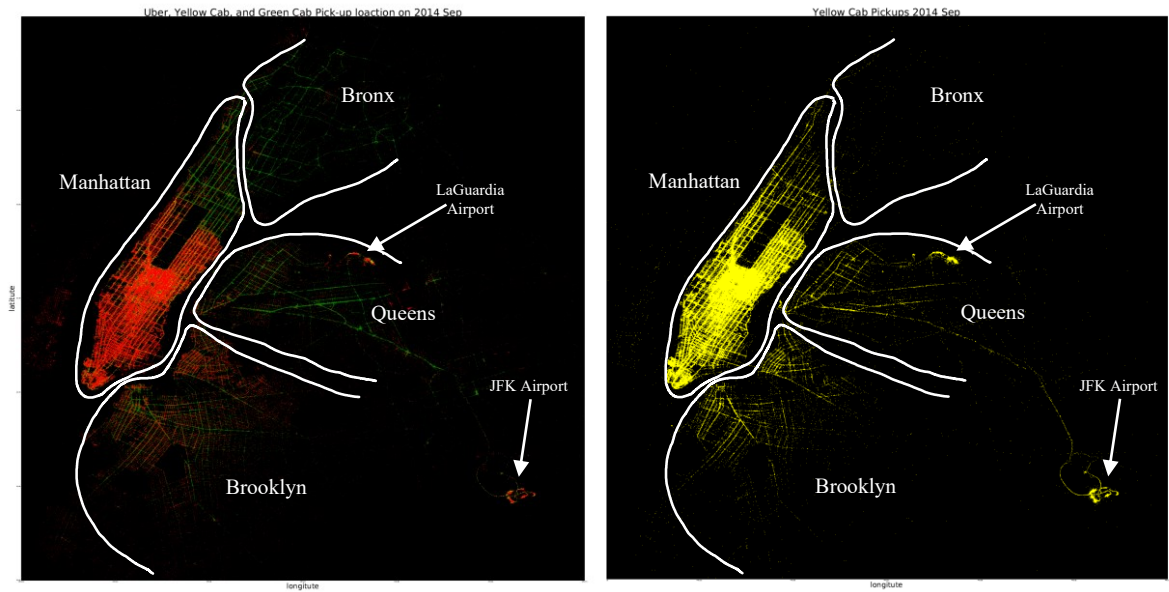

Figure 4- Pick-up locations of Yellow Cab, Green Cab, and Uber in September 2014

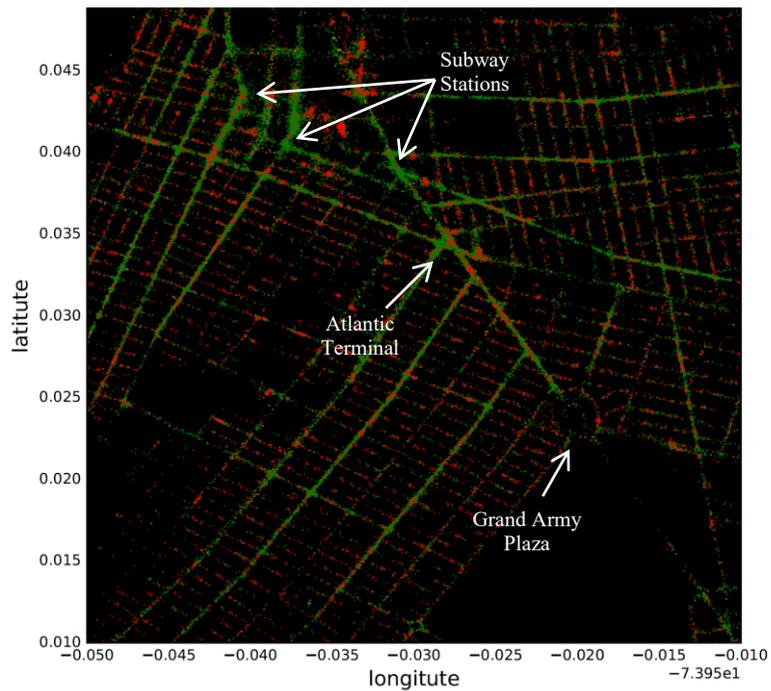

Figure 5- Passenger pickup by Yellow Cab, Green Cab, and Uber in Park Slop neighborhood, Brooklyn

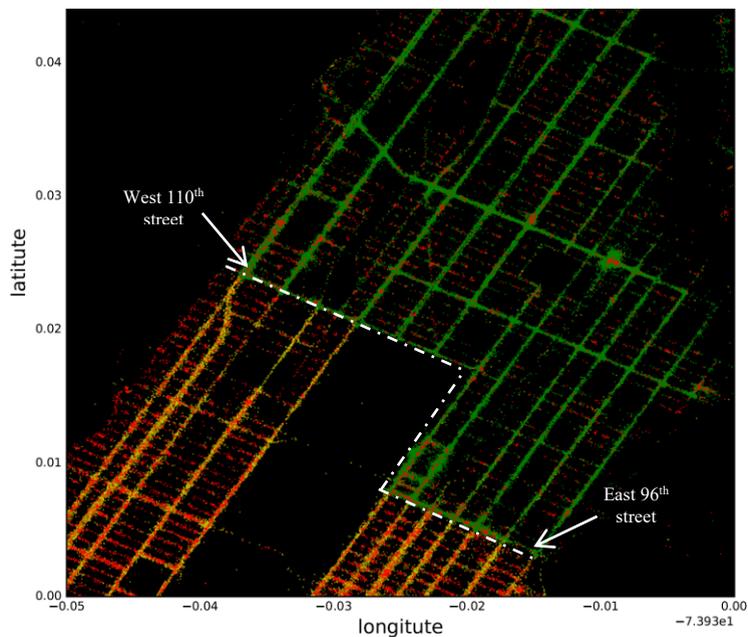

Figure 6- Yellow Cab, Green Cabs, and Uber's pickups around East 96th and West 101st streets in Upper Manhattan

Figure 7 shows the pick-up location of taxis (i.e., Uber, Green Cab, and Yellow Cab) around LaGuardia airport and Jackson Heights neighborhood. Yellow Cab and Uber pick up a large number of passengers at the airport. Green Cab is not allowed to pick up any passenger, and it is only permitted to drop off. Most of the pick-ups around the airport (i.e., Jackson Heights neighborhood, Queens) are happened by Green Cab. Green Cab's number of the pick-up in this area is much higher than any other location in Queens. Uber and Yellow Cab have a relatively small number of pick-ups in the Jackson Heights neighborhood.

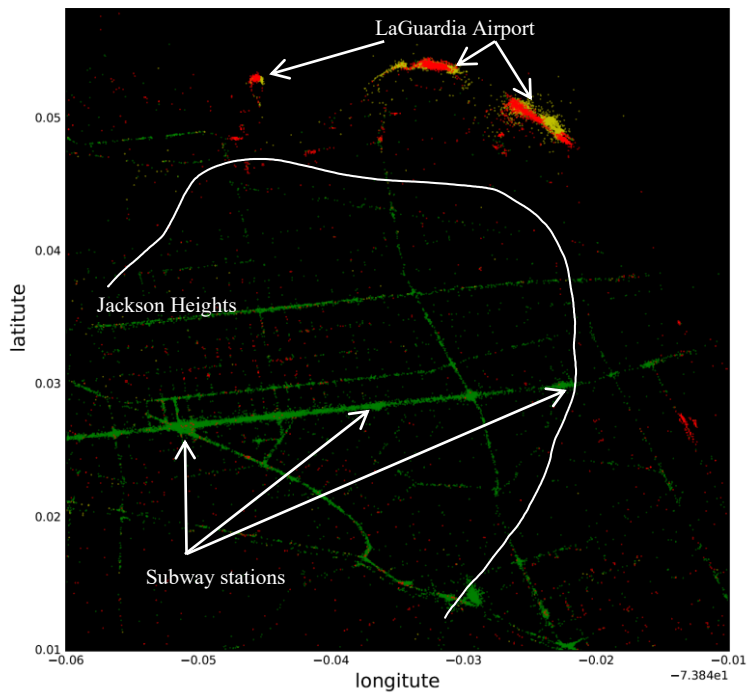

Figure 7- Taxis' pickups around LaGuardia Airport and Jackson Heights neighborhood in Queens

We have visualized change in pick-up distribution of the taxis from April to September 2014. We have identified market expansion differences between Uber and Cabs in some areas. Figure 8 shows one of these areas around the Queens Museum in April and September 2014. The left picture shows the area in April, and the right picture shows the same area in September 2014. During this period, Yellow and Green Cab's pick-ups do not show significant changes. They concentrate on Subway stations, Citi stadium, and main streets. However, Uber has expanded its pick-ups in Queens Museum, theater, and Terrace in the Park, where there has not been a market for any hailing services before (i.e., Yellow Cab or Green Cab). Uber has found its own unique market or responded change in demand around Queens Museum, beyond traditional markets (e.g., subway stations, transportation hubs, or residential areas).

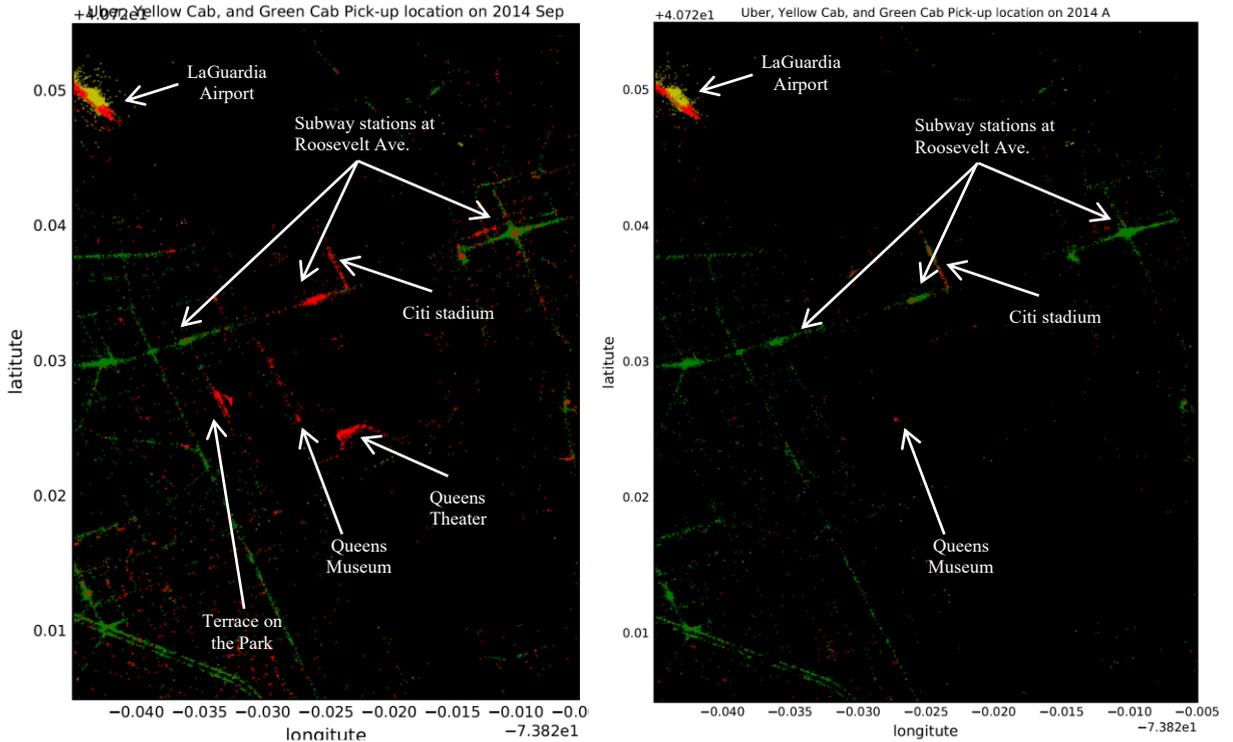

Figure 8- Taxis' pick-ups around the Queens Museum in April (on the right) and September (on the left) 2014

The multi-agent model is utilized to evaluate several hypotheses. The model is run with five agents of each type (e.g., Yellow Cab and Uber) for 90,000 iterations. In order to simulate each of the hypotheses, the general multi-agent model has been adjusted to illustrate the hypotheses' conditions and characteristics. Table 2 shows configurations of the model for each hypothesis based on types of agents, available passenger areas, and agents' regulations. In Hypothesis *A,* one high-density and one low-density passenger area are modeled. Cabs (i.e., Yellow or Green Cab) and Ubers are allowed to pick-up and drop-off passengers at any location. In Hypothesis *B,* we simulated a taxi system with one fix low and one fix high demand area, which represents current passenger demands (e.g., residential areas and transportation hubs). Also, dynamic temporary demands were considered, which were generated several times during the simulation in different locations and after a short period, the demands disappeared. These dynamic and new demand areas represent a change in taxi demand due to events, new subway stations, or new markets (e.g., Queens Museum). To simulate Hypothesis *C*, we developed a passenger demand area with equal distribution of passengers. In this case, Yellow Cabs and Ubers are allowed to pick-up and drop-off passengers at any location. However, Green Cabs are not permitted to pick up passengers in half of the area (i.e. like above and below of East 96th and West 101st streets in Upper Manhattan). We simulate Hypothesis *D* with a very high-density passenger area, representing LaGuardia airport, and a wide low-density passenger area, representing Jackson Heights. Yellow Cab and Uber are permitted to pick up passengers in both areas, but Green Cab could only pick up passengers in the low-density area.

Table 2- The Multi-agent model's configurations for each hypothesis

|  | Agent Type | Passenger distribution | Regulations |
| --- | --- | --- | --- |
| Hypothesis A | Cab (Yellow or Green) and Uber | One high-density and one low-density area | - |
| Hypothesis B | Cab (Yellow or Green) and Uber | One high-density, one low-density area, and some temporary new demand areas | - |
| Hypothesis C | Yellow Cab, Green Cab, and Uber | Equally distributed high-density passenger areas | Green Cab is not allowed to pick up on half of the total area |
| Hypothesis D | Yellow Cab, Green Cab, and Uber | One high-density and one low-density area | Green Cab is not allowed to pick up on a high-density area. |

The results of the multi-agent model for the hypotheses are presented in Figure 9. The generated new scenarios' results are shown in Figure 10.

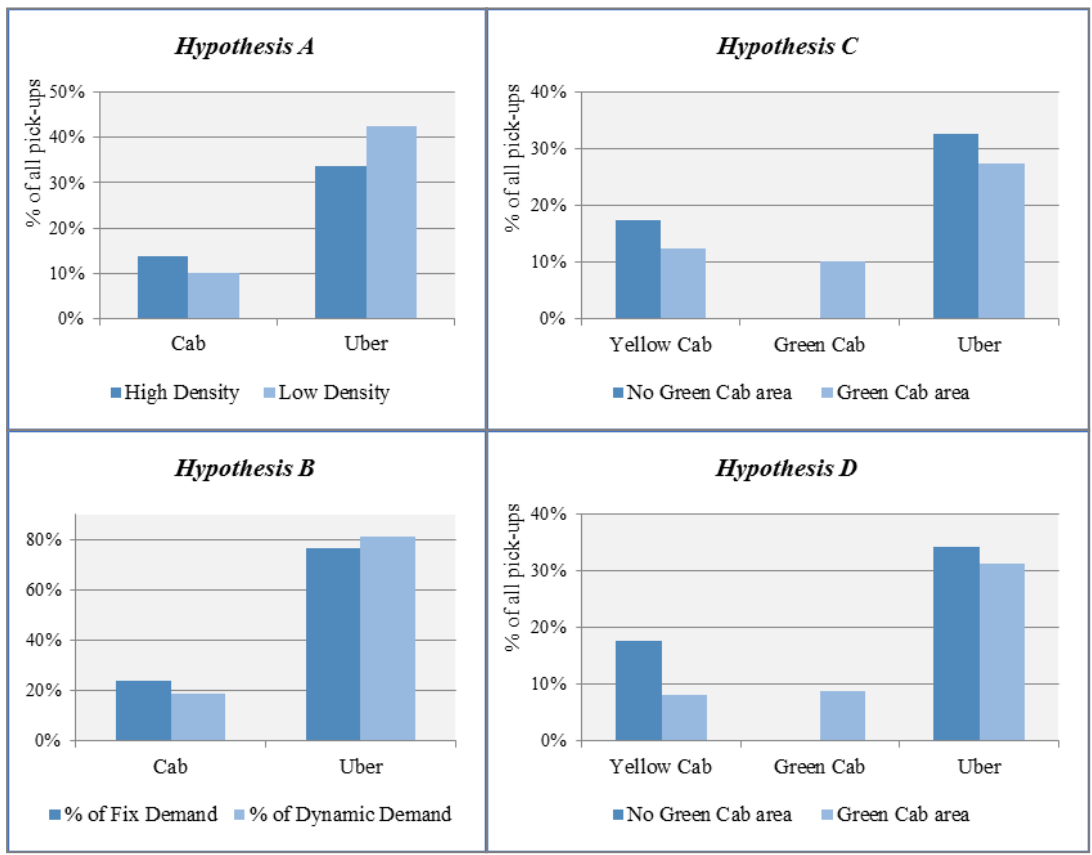

Figure 9- The hypotheses modeling results

12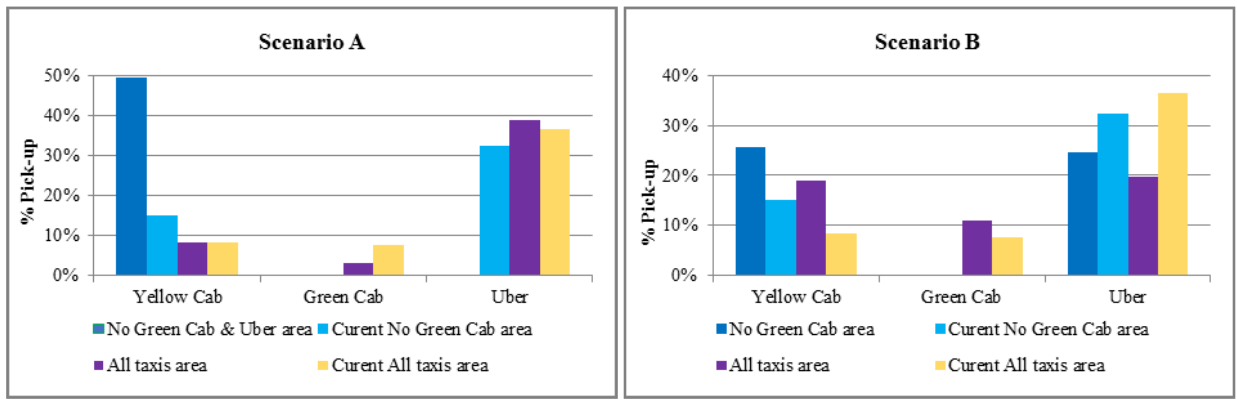

Figure 10- The scenarios' results

**References**

1. TLC. TLC 2016 Fact Book.; 2016. http://www.nyc.gov/tlcfactbook.
2. Toddwschneider. Analyzing 1.1 Billion NYC Taxi and Uber Trips, with a Vengeance. http://todd/wschneider.com/posts/analyzing-1-1-billion-nyc-taxi-and-uber-trips-with-a-vengeance. Published 2017.
3. Li, Xiaoming, Zhihan Lv, Weixi Wang, Baoyun Zhang, Jinxing Hu, Ling Yin SF. WebVRGIS based traffic analysis and visualization system. Adv Eng Softw. 2016;93:1-8.
4. Wikipedia. Boro Taxi. https://en.wikipedia.org/wiki/Boro_taxi. Published 2017.
5. Liu Y, Wang F, Xiao Y, Gao S. Urban land uses and traffic "source-sink areas": Evidence from GPS-enabled taxi data in Shanghai. Landsc Urban Plan. 2012;106(1):73-87. doi:10.1016/j.landurbplan.2012.02.012.
6. Austin, Drew PZ. Taxicabs as public transportation in Boston, Massachusetts. ransportation Res Rec J Transp Res Board. 2012;2277:65-74.
7. Haggag, Kareem, Brian McManus GP. Learning by driving: Productivity improvements by New York City taxi drivers. Am Econ J Appl Econ. 2017;9(1):70-95.
8. Dong Y. Revealing New York taxi drivers' operation patterns focusing on the revenue aspect. In: Intelligent Control and Automation (WCICA), 2016 12th World Congress On. IEEE. ; 2016.
9. Cramer, Judd ABK. Disruptive change in the taxi business: The case of Uber. Am Econ Rev. 2016;106(5):177-182.
10. Geradin D. Should Uber be allowed to compete in Europe? And if so how? 2015.
11. Peng GH, Sun DH. A dynamical model of car-following with the consideration of the multiple information of preceding cars. Phys Lett Sect A Gen At Solid State Phys. 2010;374(15-16):1694-1698. doi:10.1016/j.physleta.2010.02.020.
12. Posen HA, Information C. Ridesharing in the Sharing Economy : Should Regulators Impose Uber. 2015;1:405-433.
13. Wallsten S. The competitive effects of the sharing economy: how is Uber changing taxis? Technol Policy Inst. 2015;22.
14. Watkins CJCH. Learning from delayed rewards. 1989.
15. Cybenko, G., R. Gray KM. Q-learning: A Tutorial and Extensions. Math Neural Networks. 1997.
16. NYC Government . TLC Data. http://www.nyc.gov/html/tlc/html/technology/data.shtml.
17. Github. Import public NYC taxi and Uber trip data into PostgreSQL. https://github.com/toddwschneider/nyc-taxi-data.